\documentclass[aps,preprint,prd,showpacs,nofootinbib]{revtex4}
\usepackage{amsmath}
\usepackage{graphicx}
\usepackage{subfigure}
\usepackage{multirow}
\usepackage{bm}
\usepackage{amssymb}
\usepackage{mathtools}
\usepackage{enumerate}
\usepackage{color}
\usepackage[colorlinks,linkcolor=magenta,anchorcolor=cyan,citecolor=blue]{hyperref}
\def\be{\begin{equation}}
\def\ee{\end{equation}}
\def\ba{\begin{eqnarray}}
\def\ea{\end{eqnarray}}
\def\lf{\left}
\def\rt{\right}

\begin{document}
\title{$T_0$ censorship of early dark energy and AdS vacua}

\author{Gen Ye$^{1}$\footnote{yegen14@mails.ucas.ac.cn}}
\author{Yun-Song Piao$^{1,2,3,4}$\footnote{yspiao@ucas.ac.cn}}

\affiliation{$^1$ School of Physics, University of Chinese Academy of
    Sciences, Beijing 100049, China}

\affiliation{$^2$ Institute of Theoretical Physics, Chinese
    Academy of Sciences, P.O. Box 2735, Beijing 100190, China}

\affiliation{$^3$ School of Fundamental Physics and Mathematical
Sciences, Hangzhou Institute for Advanced Study, UCAS, Hangzhou
310024, China}

\affiliation{$^4$ International Center for Theoretical Physics
Asia-Pacific, Beijing/Hangzhou, China}

\begin{abstract}

Present-day temperature $T_0$ of cosmic microwave background has
been precisely measured by the FIRAS experiment. We
identify that the early dark energy (EDE) (non-negligible around
matter-radiation equality) scenario can remain compatible with the
FIRAS result, while lifting the Hubble constant $H_0$. We perform Monte Carlo Markov chain analysis to
confirm our observations. We also present an $\alpha$-attractor
Anti-de Sitter (AdS) model of EDE, in which the AdS
depth is consistently varied in the Monte Carlo Markov chain analysis. We found that
our datasets weakly hinted the existence of an AdS phase near
recombination with $H_0\sim 73$km/s/Mpc at 1$\sigma$ region in the
best-fit model.

\end{abstract}
\maketitle

\section{Introduction}

The Hubble constant $H_0$, the present-day expansion rate of the
Universe, sets the scale of the current Universe. Local
measurements of $H_0$ yield $H_0\gtrsim 73$km/s/Mpc
\cite{Riess:2019cxk,Freedman:2019jwv,Huang:2019yhh,Wong:2019kwg}
(e.g.the SH0ES group reports $H_0=74.03\pm1.42$km/s/Mpc
\cite{Riess:2019cxk,Riess:2020xrj}), which shows $>4\sigma$ discrepancy
\cite{Riess:2020sih} compared with the Planck result
$H_0=67.72\pm0.40$km/s/Mpc \cite{Aghanim:2018eyx}. This
discrepancy (called ``\textit{Hubble tension}") can hardly be explained by systematic errors \cite{Verde:2019ivm}.

However, the analysis of Planck is based on $\Lambda$CDM and
probes of high redshift physics, i.e. cosmic microwave
background (CMB) and baryon acoustic oscillations (BAO). Thus the
Hubble tension might be a hint of beyond-$\Lambda$CDM physics,
specially before recombination
\cite{Bernal:2016gxb,Aylor:2018drw,Knox:2019rjx,Lyu:2020lwm}. One
possibility is early dark energy (EDE)
\cite{Poulin:2018cxd,Agrawal:2019lmo,Lin:2019qug,Alexander:2019rsc,Smith:2019ihp,Sakstein:2019fmf,Niedermann:2019olb,Ye:2020btb,Braglia:2020bym,Niedermann:2020dwg}
(see also
\cite{Ballesteros:2020sik,Zumalacarregui:2020cjh,Braglia:2020iik}
for modified gravity). EDE is non-negligible only for
a short period near matter-radiation equality and before
recombination (the Universe after recombination is
$\Lambda$CDM-like), which results in a suppressed sound horizon,
and thus $H_0\gtrsim 70$km/s/Mpc.

Recently, it has been found in Ref.\cite{Ye:2020btb} that the
existence of Anti de-Sitter (AdS) vacua around recombination can
effectively lift $H_0$ to $\sim 73$km/s/Mpc at $1\sigma$ region.
The cosmologies with an AdS phase at low-$z$ have been studied in
Refs.\cite{Visinelli:2019qqu,Akarsu:2019hmw,Calderon:2020hoc}. The
AdS vacua is ubiquitous in the landscape (consisting of all
effective field theories with consistent UV-completion)
\cite{Ooguri:2006in,Obied:2018sgi}, see also
\cite{Piao:2004me,Li:2019ipk} for inflation with multiple AdS
vacua. The AdS potential in Ref.\cite{Ye:2020btb} is only a
phenomenological one with the AdS depth being fixed
by hand rather than varied in the analysis. Thus it is
significant to search for AdS-EDE models originating
from UV-complete theories in the cosmological setup with varying AdS depth in the Monte Carlo Markov Chain (MCMC) analysis .

Precise measurement of the present-day CMB $T_0$ from the
COBE/FIRAS experiment, independent of Planck, yields
\cite{Fixsen:1996nj,Fixsen:2009ug} \be
T_{0,FIRAS}=2.72548\pm0.00057K. \label{COBET0}\ee Based on
$\Lambda$CDM, the Planck and BAO data yields $T_0=2.718\pm
0.021K$ \cite{Ade:2015xua}, consistent with $T_{0,FIRAS}$.
However, the $T_0$ deduced from the Planck and SH0ES data, assuming $\Lambda$CDM, has $>4\sigma$
discrepancy compared with $T_{0,FIRAS}$, called $T_0$ tension in
Ref.\cite{Ivanov:2020mfr}, see also
\cite{Bose:2020cjb,Bengaly:2020vly} for recent studies. This might
be yet another hint of new physics beyond $\Lambda$CDM.

In this paper, we identify, at the cosmological
parameter level, how the EDE scenario lifts $H_0$, while staying
compatible with $T_{0,FIRAS}$, at the cost of a parameter shift to
a larger $\omega_m$. We perform MCMC
analysis to confirm our observations. We also present
a theoretically well-motivated AdS-EDE model as well as the
corresponding MCMC analysis with the AdS depth consistently
varied. It is noticed that the full datasets weakly hinted
the existence of an AdS phase near recombination. Low-$z$
resolutions to the Hubble tension have also been discussed, see
e.g.\cite{Vagnozzi:2019ezj,Benevento:2020fev,Haridasu:2020xaa,DiValentino:2020kha}
for different perspectives. As a contrast, we also show that
$w$CDM models with a constant equation of state parameter
$w\lesssim -1.3$ of dark energy at low-$z$ seem incompatible with
$T_{0,FIRAS}$. Throughout this paper we assume a spatially flat
Universe.

\section{Early dark energy and AdS}\label{sec:simulation}

EDE may be non-negligible only for a short epoch decades before
recombination \cite{Poulin:2018cxd,Agrawal:2019lmo}. The injection
of EDE energy results in a larger Hubble rate $H(z\gtrsim
z_{rec})$ prior to recombination, so a suppressed sound horizon
$r_s=\int_{z_{rec}}^{\infty}dz/H(z)$. The spacing of CMB acoustic
peaks perfectly sets the angular scale $\theta_{CMB}$, \be
\theta_{CMB}=\frac{r_s(z_{rec})}{D_A(z_{rec})},\label{thetas}\ee
where \begin{equation} D_A(z_{rec})\equiv\int^{z_{rec}}_0
\frac{dz}{H(z)}=\frac{1}{T_0}\int^{T_{rec}}_{T_0}\frac{dT}{H(T)}
\label{DA}\end{equation} and $z_{rec}\sim 1100$ is the
recombination redshift. $D_A(z_{rec})$ is the comoving angular
distance, which is sensitive only to post recombination physics.
Generally, $D_A$ is anti-correlated with $H_0$, so for constant $\theta_{CMB}$,
$H_0\sim r_s^{-1}$ will increase.

In the AdS-EDE model \cite{Ye:2020btb}, initially the scalar field
sits at the hillside of its potential $V(\phi)$, and $\rho_\phi$
is negligible. It will roll down the potential sometime near matter-radiation equality (when $\rho_\phi/\rho_{tot}\sim 10\%$), and
roll into an AdS phase. In the AdS region, we have
$w_\phi=p_{\phi}/\rho_{\phi}>1$, so that $\rho_\phi\sim
a^{-3(1+w)}$ will more quickly redshift away (in
Refs.\cite{Poulin:2018cxd,Agrawal:2019lmo,Smith:2019ihp} the
dissipation of $\rho_\phi$ is less effective by oscillation with
cycle-averaged $w<1$, see also
Refs.\cite{Niedermann:2019olb,Niedermann:2020dwg} for different
mechanisms). This is crucial for having a larger injection of
$\rho_\phi$ ($>10\%$), thus a higher $H_0$. $\rho_\phi$
injected must be dissipated rapidly enough so that it is negligible around
recombination, or it will interfere with the fit of $\Lambda$CDM
to CMB data. After that, the field will climb up to the $\Lambda>0$
region, and the Universe is settled to be $\Lambda$CDM-like until
now.

\begin{figure}[h]
\centering
\includegraphics[width=0.6\linewidth]{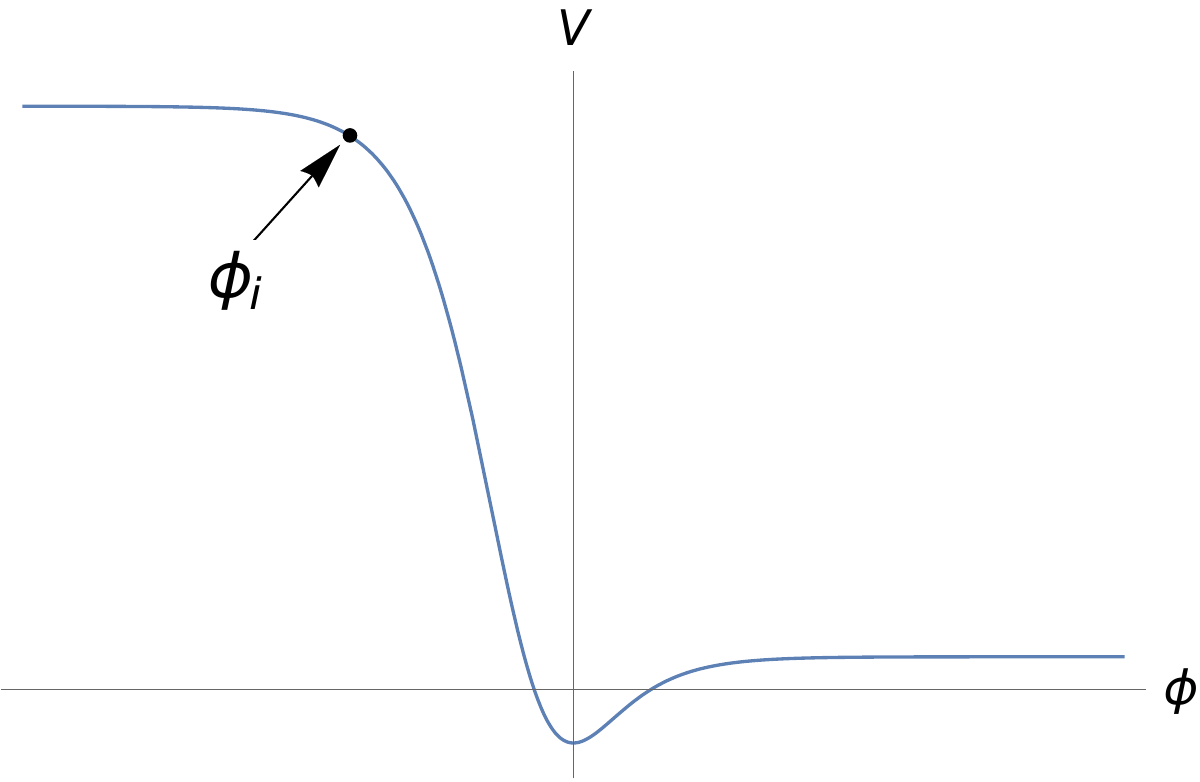}
\caption{Potential \eqref{sg potential}, plotted only for
illustration. The scalar field initially sits at $\phi_i$ near the
high plateau. It begins rolling down the potential around matter-radiation equality,
passing through the AdS region near $\phi\simeq 0$ and finally
climbs up the low plateau responsible for the current dark energy.}
\label{sg potential plt}
\end{figure}

The potential $V(\phi)$ in Ref.\cite{Ye:2020btb} is only a
phenomenological one constructed by gluing a $\phi^4$ like AdS minima with a late cosmological constant phase. Inspired by
the $\alpha$-attractor \cite{Carrasco:2015rva,Akrami:2017cir}, we take $V(\phi)$ as (see Fig-\ref{sg potential plt})
\begin{equation}\label{sg potential}
V(\phi)=V_0\left[1-\exp\left(-\gamma\tanh(\frac{\phi}{M_p\sqrt{6\alpha}})\right)\right]^2-V_0+V_\Lambda.
\end{equation}
For $\phi\ll -M_p(6\alpha)^{1/2}$, we have a high plateau
$V(\phi)\sim e^{2\gamma}V_0$ responsible for EDE. For $\phi\gg
M_p(6\alpha)^{1/2}$, $V(\phi)=V_\Lambda$ behaves like a cosmological constant in the current Universe. In Ref.\cite{Akrami:2017cir}, the high
plateau drives inflation in the early Universe, in which case
$\gamma=\ln({H_{inf}\over H_\Lambda})\gg 1$.

Here, the AdS-EDE model with potential \eqref{sg potential} will be briefly
called $\alpha$AdS. Initially, $\rho_{\phi_i}=V(\phi_i)\simeq (0.1$eV$)^4$, roughly equal to height of
the high plateau $e^{2\gamma}V_0$ if $\alpha\ll 1$. In the MCMC analysis, we choose ${6\alpha}=(0.15)^2\ll1$ for simplicity, thus only $V_0, \
\gamma, \ V_\Lambda$ are free parameters. The minima of
potential \eqref{sg potential} is $V_{min}=-V_0+V_\Lambda$ at $\phi=0$. Whether potential \eqref{sg potential} accommodates AdS vacua or not depends on the value of $\gamma$. The
existence of an AdS phase requires $V_0\gtrsim V_{\Lambda}$, i.e.\be
\gamma\lesssim {1\over 2}\ln{V(\phi_i)\over V_\Lambda} \simeq 13,
\label{gamma13}\ee where $V_\Lambda\sim (10^{-4}$eV$)^{4}$ is the
current dark energy scale. In the limit of large $\gamma$,
the $\alpha$AdS model reduces to a run-away model
\cite{Lin:2019qug,Alexander:2019rsc} with $V(\phi>0)\sim
V_\Lambda$.

\section{$T_0$ censorship of beyond-$\Lambda$CDM models}\label{sec:result}

\subsection{Dataset }
Our datasets consist of the Planck18 high-$l$ and low-$l$
TT,EE,TE and lensing likelihoods (P18) \cite{Aghanim:2018eyx}, the
BOSS DR12 \cite{Alam:2016hwk} with its full covariant matrix for
BAO measurements as well as the 6dFGS \cite{Beutler:2011hx} and MGS of SDSS
\cite{Ross:2014qpa} for low-$z$ BAO, and the Pantheon data (SN) \cite{Scolnic:2017caz}.
Recent SH0ES result $H_0=74.03\pm 1.42$km/s/Mpc
\cite{Riess:2019cxk} is employed as a Gaussian prior ($H_0$). We
modified the Montepython-3.3
\cite{Audren:2012wb,Brinckmann:2018cvx} and CLASS
\cite{Lesgourgues:2011re,Blas:2011rf} codes to perform the MCMC
analysis.

\begin{table}
\centering
\begin{tabular}{|c|c|c|c|c|c|c|}
\hline
&$100\omega_b\hat{T}^{-3}$&$\omega_{cdm}\hat{T}^{-3}$&$H_0$&$T_0$&$w$&$r_s$\\
\hline
$\Lambda$CDM+P18&$2.195_{-0.018}^{+0.016}$&$0.1186_{-0.0016}^{+0.0016}$&$69.2_{-2.3}^{+2.2}$&$2.661_{-0.06}^{+0.059}$&$-1$&$148.4_{-3.7}^{+3.3}$\\
\hline
$\Lambda$CDM+P18+BAO&$2.189_{-0.015}^{+0.014}$&$0.1194_{-0.0011}^{+0.0011}$&$67.68_{-0.5}^{+0.5}$&$2.701_{-0.016}^{+0.016}$&$-1$&$146_{-0.78}^{+0.8}$\\
\hline  \hline
$\phi^4$&$2.227_{-0.019}^{+0.019}$&$0.1285_{-0.0042}^{+0.0043}$&$70.94_{-1.1}^{+1}$&$2.709_{-0.016}^{+0.015}$&$-1$&$140.5_{-2.3}^{+2.1}$\\
\hline
$\phi^4$AdS&$2.296_{-0.018}^{+0.017}$&$0.1344_{-0.0022}^{+0.0019}$&$72.6_{-0.6}^{+0.53}$&$2.716_{-0.015}^{+0.016}$&$-1$&$136.9_{-0.91}^{+1.1}$\\
\hline
$\alpha$AdS&$2.273_{-0.047}^{+0.044}$&$0.1345_{-0.0025}^{+0.002}$&$72.57_{-0.53}^{+0.52}$&$2.709_{-0.016}^{+0.015}$&$-1$&$136.8_{-0.78}^{+0.95}$\\
\hline\hline
$w$CDM+P18&$2.191_{-0.017}^{+0.017}$&$0.1192_{-0.0016}^{+0.0015}$&$74.42_{-1.8}^{+5.6}$&$2.715_{-0.071}^{+0.064}$&$-1.244_{-0.16}^{+0.2}$&$145.3_{-3.9}^{+3.8}$\\
\hline
$w$CDM+P18+$\theta_{BAO}^{\perp}$+$H_0$&$2.184_{-0.015}^{+0.015}$&$0.1201_{-0.0012}^{+0.0011}$&$74.01_{-1.5}^{+1.4}$&$2.77_{-0.025}^{+0.027}$&$-1.322_{-0.084}^{+0.095}$&$142.2_{-1.4}^{+1.3}$\\
\hline
\end{tabular}
\caption{Mean and $1\sigma$ results of all the chains. All EDE
models ($\phi^4$ \cite{Agrawal:2019lmo}, $\phi^4$AdS
\cite{Ye:2020btb}, $\alpha$AdS) are confronted with
P18+BAO+SN+$H_0$ datasets.} \label{mcmc table}
\end{table}

Here, we regard $T_0$ as an MCMC parameter. We sample the
cosmological parameter set $\{\hat{T}_0^{-3}\omega_b,
\hat{T}_0^{-3}\omega_{cdm}, H_0,
\ln(10^{10}A_s\hat{T}_0^{1+n_s}),n_s,\tau_{reio},T_0\}$ for
$\Lambda$CDM, where $\hat{T}_0\equiv T_0/T_{0,FIRAS}$ and
$\bar{\omega}_{b/cdm}T_{0,FIRAS}^3\equiv\hat{T}_0^{-3}\omega_{b/cdm}$
(reducing degeneracy between $H_0$, $\omega_{b/cdm}$ and $T_0$,
see Ref.\cite{Ivanov:2020mfr}). The $w$CDM models introduce one
more MCMC parameter $w$. Beyond that, the EDE-like models have
additional parameters $\{\omega_{scf},\ln(1+z_c)\}$. As described
in Refs.\cite{Poulin:2018cxd,Agrawal:2019lmo,Ye:2020btb}, $z_c$ is
the redshift at which the field $\phi$ starts rolling and
$\omega_{scf}=\rho_{\phi}/\rho_{tot}$ is the energy fraction of
EDE at $z_c$. Moreover, the $\alpha$AdS model \eqref{sg potential}
has yet a parameter $\gamma$. Note that by varying
$\gamma$, and thus the AdS depth (which was fixed by
hand in Ref.\cite{Ye:2020btb}), we can explore both the AdS and
non-AdS potentials in one MCMC run. Once
$\{\omega_{scf},\ln(1+z_c),\gamma\}$ are fixed, $V_\Lambda$ will
be set by matching the budget equation
$\Omega_{DE}=1-\Omega_m-\Omega_r$. The field initially sits around
the high plateau $3\omega_{scf}M_p^2H^2(z_c)\sim e^{2\gamma}V_0$,
so the minimal value $V_{min}$ of potential \eqref{sg potential}
\be V_{min}\sim -3\omega_{scf}M_p^2H^2(z_c)e^{-2\gamma}+V_\Lambda
\label{Vmin}\ee is roughly set by $\gamma$, $\omega_{scf}$ and
$z_c$. When $\gamma\lesssim 13$, $V_{min}<0$ is AdS-like, see
\eqref{gamma13}.

\subsection{Physical consideration}\label{sec:constraint}

In our dataset, CMB and BAO play significant roles. Thus it
is worthwhile to highlight their constraints on parameters
$\{h_0,T_0,|w|,\bar{\omega}_m\}$, where
$h_0=H_0\times(100$km/s/Mpc$)^{-1}$, which helps to clarify the
MCMC results in Sect-\ref{sec:MCMC}.

We assume a spatially-flat Universe, which is $w$CDM-like after
recombination. We can Taylor expand $D_A(z_{rec})$ around a
best-fit Planck $\Lambda$CDM model (by performing partial
derivatives with respect to one of
$\{h_0,T_0,|w|,\bar{\omega}_m\}$) to estimate its dependence on
$\{h_0,T_0,|w|,\bar{\omega}_m\}$. Using $\Omega_m\simeq 0.3$ and
$\Omega_{DE}\simeq 0.7$, for fixed $\theta_{CMB}$ in
\eqref{thetas}, we have
\begin{equation}\label{CMB}
\quad (r_s T_0) h_0^{0.19} T_0^{0.21} |w|^{-0.09}
\bar{\omega}_m^{0.4}=const.
\end{equation}

The BOSS experiment \cite{Alam:2016hwk} sets the BAO angular scales
as \ba \theta^{\parallel}_{BAO}=r_d
H(z_{eff})/(1+z_{eff}),\quad\quad
\theta^{\perp}_{BAO}=\frac{r_d}{D_A(z_{eff})},\ea where $z_{eff}$
is the effective redshift bins of BOSS DR12 data
(i.e. $z_{eff}=0.38,0.51,0.61$ \cite{Alam:2016hwk}), and $r_d$ is
the comoving sound horizon at the baryon drag epoch.
%Unlike CMB, the integral $D_A(z_{eff})$ has a fixed
%upper limit $z_{eff}$ while $T$ at $z_{eff}$ is dependent on
%$T_0$.
Here, we take $z_{eff}=0.61$ (the results at different $z_{eff}$
only exhibit slight difference). And for fixed
$\theta_{BAO}^{\parallel}$ and $\theta_{BAO}^{\perp}$, we have
\begin{eqnarray}
\theta_{BAO}^{\parallel}:& \quad (r_d T_0) h_0^{0.51} T_0^{-0.27} |w|^{-0.26}\bar{\omega}_m^{0.24}&=const.\label{BAO-trans}\\
\theta_{BAO}^{\perp}:&\quad (r_dT_0) h_0^{0.75} T_0^{-0.63}
|w|^{-0.17} \bar{\omega}_m^{0.12}&=const.\label{BAO-rad}
\end{eqnarray}

\subsection{$T_0$-$H_0$ in MCMC results}\label{sec:MCMC}

\begin{figure}[h]
    \centering
    \includegraphics[width=0.8\linewidth]{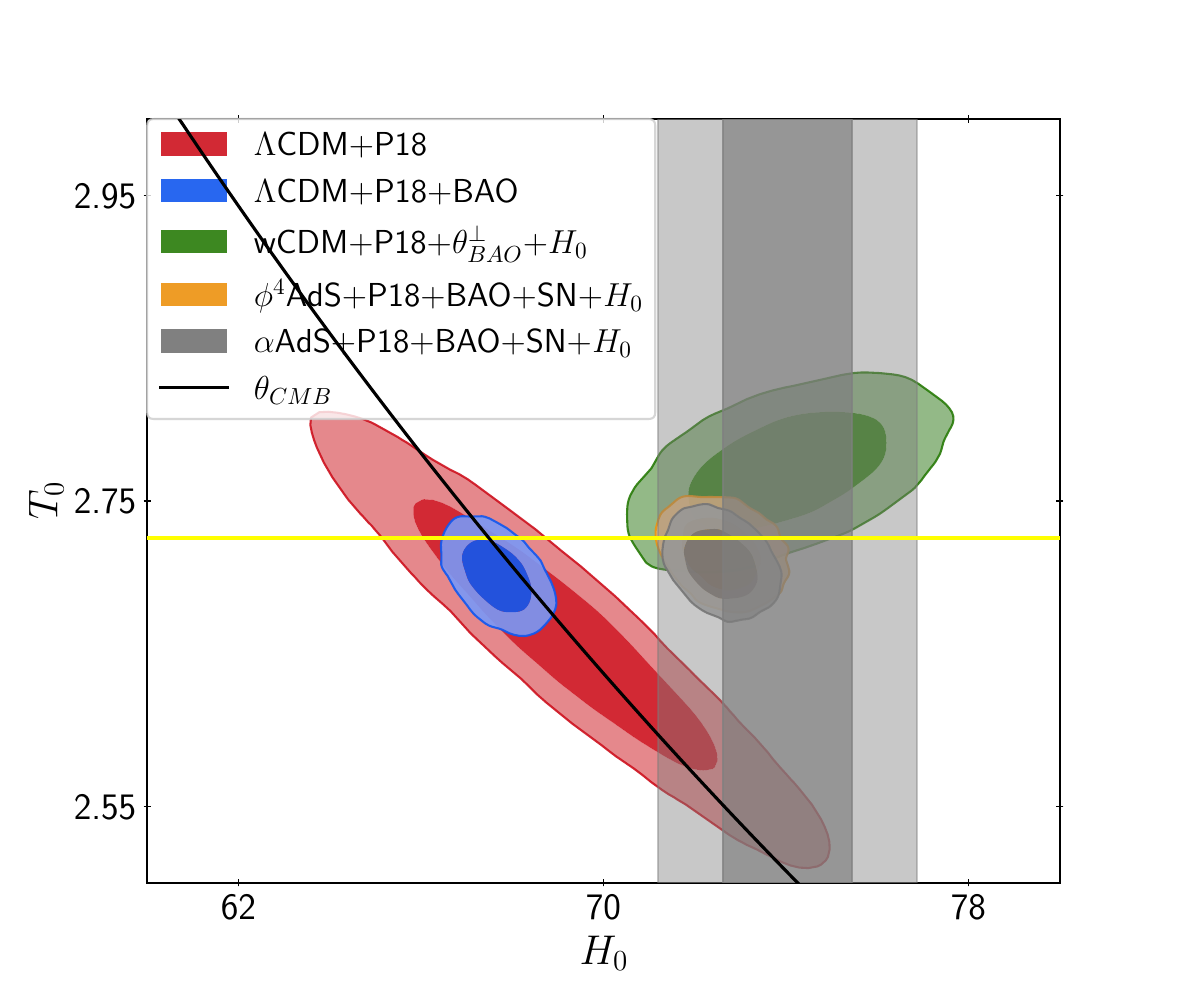}
\caption{Marginalized $1\sigma$ and $2\sigma$ contours in the
$T_0$-$H_0$ plane. The gray band is the $1\sigma$ and $2\sigma$ SH0ES result
$H_0=74.03\pm1.42$km/s/Mpc \cite{Riess:2019cxk}. The thick yellow
line depicts the FIRAS $1\sigma$ region \eqref{COBET0}
\cite{Fixsen:1996nj,Fixsen:2009ug}. Only the EDE models simultaneously lift $H_0$ and remain compatible with $T_{0,FIRAS}$.}
    \label{T0-H0}
\end{figure}

Table-\ref{mcmc table} presents the MCMC results for $\Lambda$CDM
and beyond-$\Lambda$CDM models, see also the corresponding
$T_0$-$H_0$ contours in Fig-\ref{T0-H0}. In
Appendix-\ref{apx:alpha-attractor}, we also focus on the
$\alpha$AdS model, and present the posterior distributions and marginalized contours of all the cosmological
parameters and the best-fit $\chi^2$ values per experiment.  As
expected, the existence of an AdS phase near recombination can effectively lift $H_0$
to $\sim 73$km/s/Mpc at 1$\sigma$ region.

It is well-known that certain cosmological parameters in EDE models show notable shift from those of the concordance $\Lambda$CDM model. Such parameters may receive additional constraints from the inclusion of new datasets not only depending on $H_0$. To further verify the robustness of treating local $H_0$ measurement as a simple prior, in Appendix-\ref{apx:strong lensing}, we replace the SH0ES prior with the full likelihood code\footnote{written by Stefan Taubenberger and Sherry Suyu, available at \url{https://zenodo.org/record/3632967\#.X2nuiUBuJ9B}} from the H0LiCOW group, which constrains the time delay distance $D_{\Delta t}$ and/or angular diameter distance to lens $D_d$ of six strong gravitational lenses. We confirm that using the full likelihood yields results nearly identical to those using the Gaussian prior. 

In Fig-\ref{T0-H0}, we see that the $\Lambda$CDM+P18 contour
respects Eq.\eqref{CMB} (the $\theta_{CMB}$ line). The
$\Lambda$CDM+P18 contour intersects with the SH0ES band at
$T_0\sim 2.6K$, which is inconsistent with $T_{0,FIRAS}$. As has
been pointed out in Ref.\cite{Ivanov:2020mfr}, $T_0$ yielded by
the Planck and SH0ES data has $>4\sigma$ discrepancy compared with
$T_{0,FIRAS}$.

\begin{figure}[h]
\centering
\includegraphics[width=0.8\linewidth]{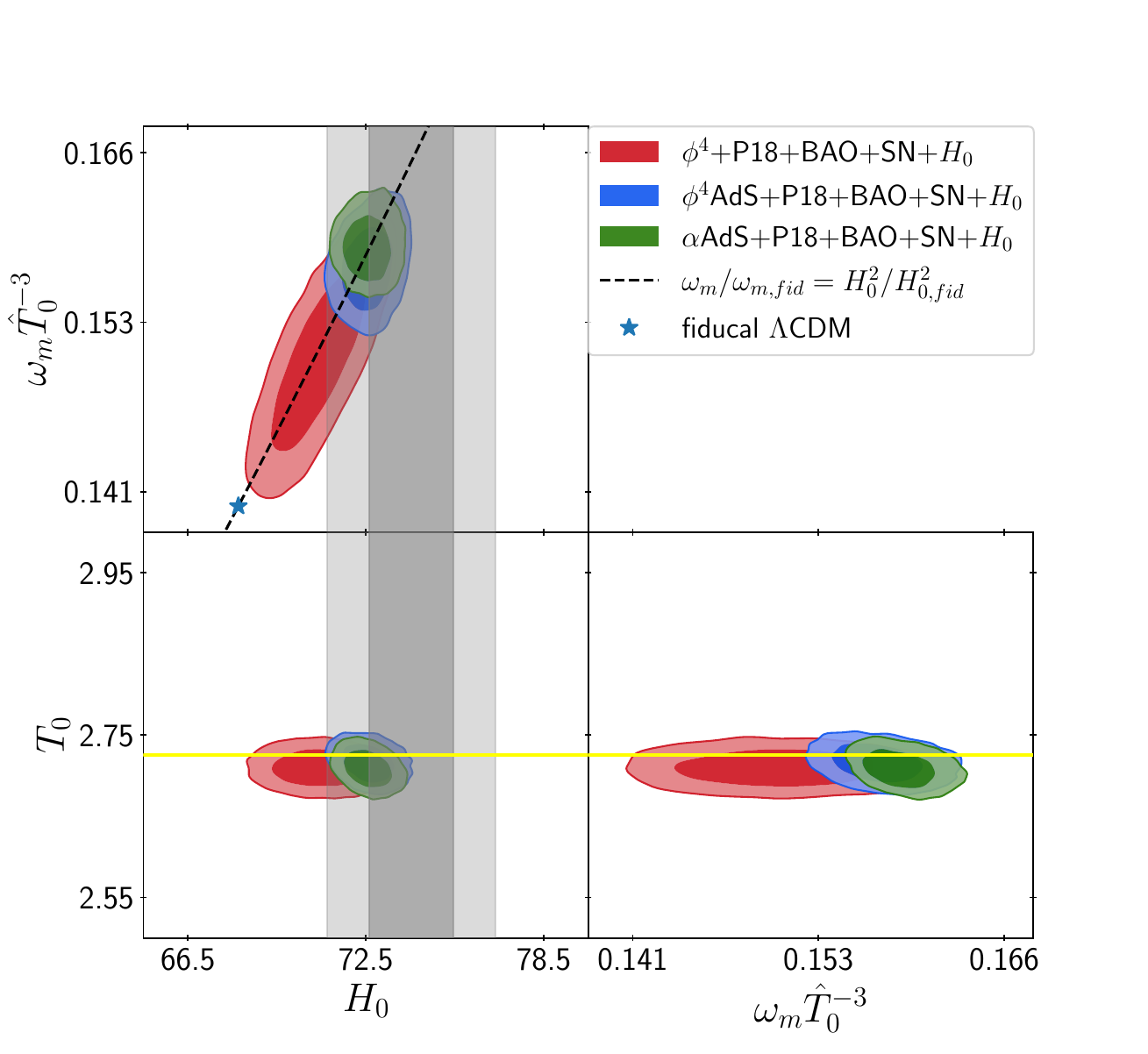}
\caption{Marginalized $1\sigma$ and $2\sigma$ contours of the EDE
models in the $\{T_0-{\bar\omega}_m-H_0\}$ space. $T_{0,FIRAS}$
and $H_0$ are plotted as described in Fig-\ref{T0-H0}. The
${\omega}_m\hat{T}_0^{-3}$-$H_0$ contours of all EDE models
respect Eq.\eqref{ede constraint} (dashed line).}
\label{ede triangle}
\end{figure}

However, the EDE scenario not only lifts $H_0$, but also is
compatible with $T_{0,FIRAS}$. This can be explained as follows.
In CMB and BAO constraints \eqref{CMB}, \eqref{BAO-trans} and
\eqref{BAO-rad}, we have $|w|= 1$ for EDE scenarios. The
Universe after recombination is $\Lambda$CDM-like, and $r_d\sim
r_s$, since the physics at and after recombination must not be affected by
EDE. Thus we (approximately) solve Eqs.\eqref{CMB}, \eqref{BAO-trans} and
\eqref{BAO-rad} for $T_0=T_{0,FIRAS}$, and have
\begin{equation}
r_sh_0\simeq const., \quad {\bar{\omega}_m}^{-1} h_0^{2}\simeq
const. \label{solution1}\end{equation} Thus though $h_0$ is lifted
due to $h_0\sim r_s^{-1}$ (essence of the EDE idea), $T_0=T_{0,FIRAS}$ needs not to be shifted.
%Thus EDE is
%compatible with $T_{0,FIRAS}$.
The expense of compatibility with $T_{0,FIRAS}$ is that \be {\bar
\omega}_{m}=\lf({h_0^2\over h^2_{0,\Lambda}}\rt){\bar
\omega}_{m,\Lambda CDM} \label{ede constraint}\ee must be
magnified. According to \eqref{ede constraint}, we actually have
$\Omega_m\simeq const$ (equivalently $\Omega_m\simeq
\Omega_{m,\Lambda CDM}$), since $\omega_m=\Omega_m h_0^2$. It is worth mentioning that the $\Omega_m\sim const.$
constraint is applicable to any mechanism that modifies the sound
horizon around recombination with a $\Lambda$CDM-like universe
after that, not necessarily EDE. As a consistency check of
\eqref{ede constraint}, for $h_{0,\Lambda CDM}\sim 0.68$ and
${\bar \omega}_{m,\Lambda CDM}\sim 0.14$ in $\Lambda$CDM (see
Table-\ref{mcmc table}), we will have ${\bar \omega}_{m}\sim 0.16$
in AdS-EDE models ($h_0\sim 0.73$), consistent with the results in
Table-\ref{mcmc table}. We plot contours of $\{H_0, T_0,
{\bar\omega}_{m}\}$ in Fig-\ref{ede triangle}. As expected, $H_0$
is lifted respecting Eq.\eqref{ede constraint}.

In $\Lambda$CDM, $\omega_m$ is difficult to adjust since it is
well constrained by Planck, but in EDE $\omega_m$ can be consistently
tuned due to the scalar field perturbations, see
Appendix-\ref{apx:omegam in ede}. This seems to cause a slightly larger
$\sigma_8$, so-called $S_8$ tension, e.g.\cite{Raveri:2018wln},
see also \cite{Hill:2020osr,Ivanov:2020ril,DAmico:2020ods}.
However, this tension is also present in $\Lambda$CDM with $\sim
2\sigma$ significance (inherited but not significantly exacerbated
in EDE, as argued in \cite{Niedermann:2020dwg,Klypin:2020tud}),
which might be related with systematic error or possible intrinsic
inconsistency of Planck data itself
\cite{Wu:2020nxz,Chudaykin:2020acu}.

The low-$z$ resolutions beyond $\Lambda$CDM have been also studied
in
e.g.\cite{DiValentino:2016hlg,DiValentino:2017iww,DiValentino:2017zyq,Yan:2019gbw,DiValentino:2019ffd,Akarsu:2019hmw,Yang:2020zuk,Yang:2020myd,Benaoum:2020qsi}.
It is usually thought that $w$CDM models with $w\simeq -1.3$ might
resolve the Hubble tension,
e.g.\cite{DiValentino:2016hlg,Vagnozzi:2019ezj,Alestas:2020mvb},
though it is disfavored by the full BAO data. However, in
Fig-\ref{T0-H0}, we see that such a solution seems also
incompatible with $T_{0,FIRAS}$.

\begin{figure}[h]
    \centering
    \includegraphics[width=0.8\linewidth]{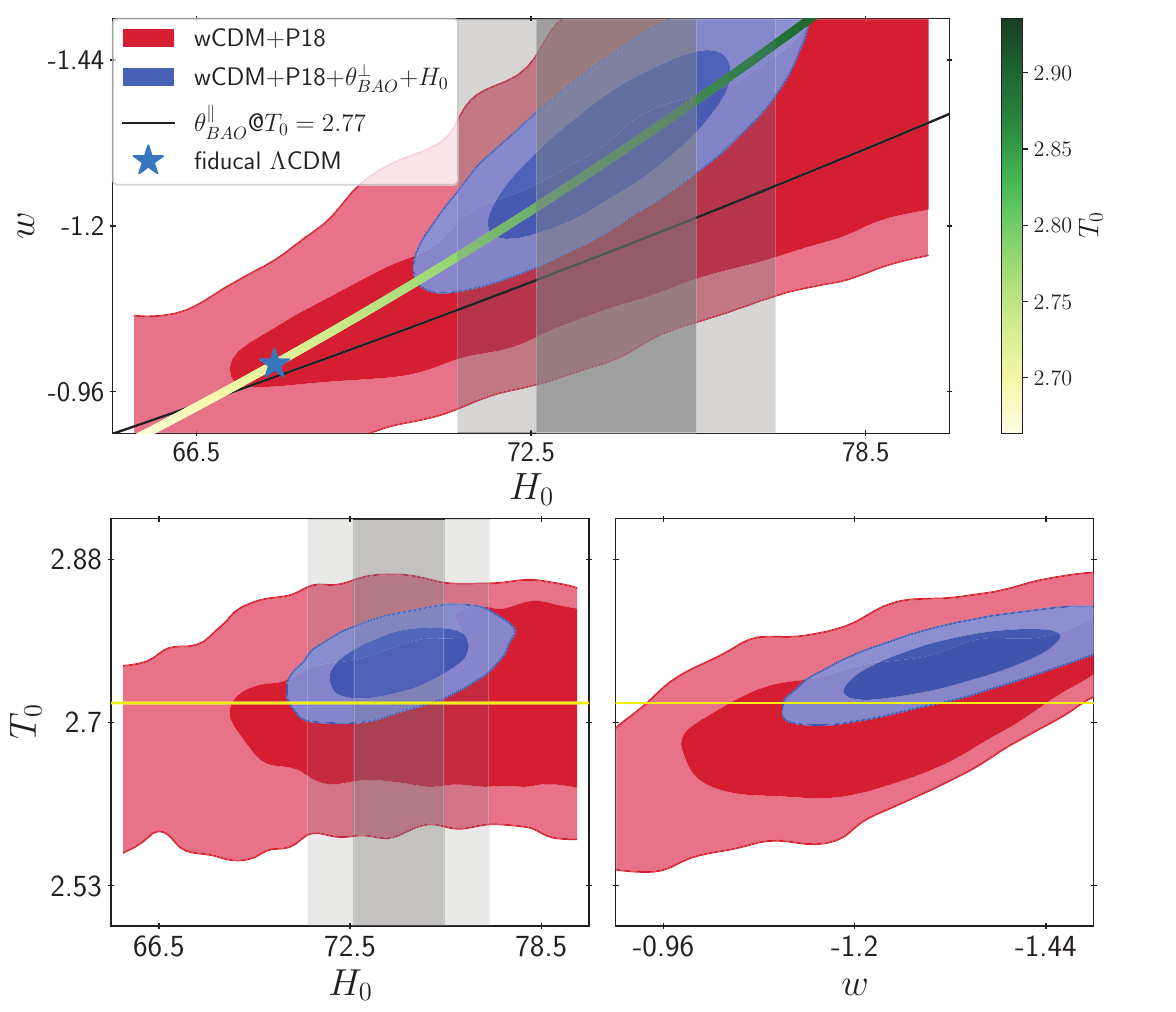}
\caption{Marginalized $1\sigma$ and $2\sigma$ contours of the
$w$CDM model in the $\{w$-$T_0$-$H_0\}$ space. $T_{0,FIRAS}$ and
$H_0$ are plotted as described in Fig-\ref{T0-H0}. Upper panel:
The rainbow line plots compatible intersections of \eqref{CMB} and
\eqref{BAO-rad} at different $T_0$, with a color coding for $T_0$.
As expected, the contour of the $w$CDM model spreads along the
predicted line. The black line plots the
$\theta_{BAO}^{\parallel}$ constraint \eqref{BAO-trans} at
$T_0=2.77K$ (see Table-\ref{mcmc table}), which suggests that
$w$CDM with $w\lesssim -1.3$ is not actually favored by BAO data.
Lower Panel: In addition, such a $w$CDM model is also inconsistent
with $T_{0,FIRAS}$.}
    \label{w-H0}
\end{figure}

The $w$CDM model, like $\Lambda$CDM, does not alter the physics
around and before recombination, so $r_sT_0$ is constant
\cite{Ivanov:2020mfr}. It is well-known that $w$CDM with $w<-1$ is
not supported by the full BAO data, e.g.recent
Ref.\cite{Alestas:2020mvb}, so we only solve Eqs.\eqref{CMB} and
\eqref{BAO-rad}, and have  \be h_0^{-3}|w|\simeq
const.,\quad T_0^{-8}|w|\simeq const.\label{solution2}\ee Note
\eqref{solution2} is conflicted with BAO constraint
\eqref{BAO-trans}, see the black line in Fig-\ref{w-H0}. Here, if
$|w|>1$, $h_0\propto |w|^{1/3}$ will be lifted. However, $T_0
\propto |w|^{1/8}$ must also be magnified, which will make $T_0$
inconsistent with the result \eqref{COBET0} of $T_{0,FIRAS}$.
Though we can fix $T_0=T_{0,FIRAS}$, and have $h_0\sim |w|^{9/19}$
for the CMB constraint \eqref{CMB}, it is obviously conflicted
with BAO constraints \eqref{BAO-trans} and \eqref{BAO-rad}. As a
consistency check of \eqref{solution2}, for $h_{0}\sim 0.68$ in
$\Lambda$CDM, we will have $w\simeq -1.3$ in $w$CDM ($h_0\sim
0.74$) but \be T_0\simeq T_{0,FIRAS}|w|^{1/8}\sim
2.8K,\label{T0w}\ee which is consistent with the $w$CDM results in
Table-\ref{mcmc table}. Here, we confront $w$CDM with P18 and
perpendicular BAO data ($\theta_{BAO}^{\perp}$). The
contours of $\{H_0, T_0, w\}$ is plotted in
Fig-\ref{w-H0}, which clearly shows the inconsistency of $w$CDM
with $T_{0,FIRAS}$. As expected, $T_0$ is lifted respecting
Eq.\eqref{T0w}.

\section{Conclusion}

It is well-known that $H_0$ and $T_0$ are basic cosmological
parameters (specially not dimensionless). Precisely measured
value $T_{0,FIRAS}$ of $T_0$ can be regarded as a censorship of
beyond-$\Lambda$CDM models resolving the $H_0$ tension.

\begin{figure}[h]
\includegraphics[width=0.75\linewidth]{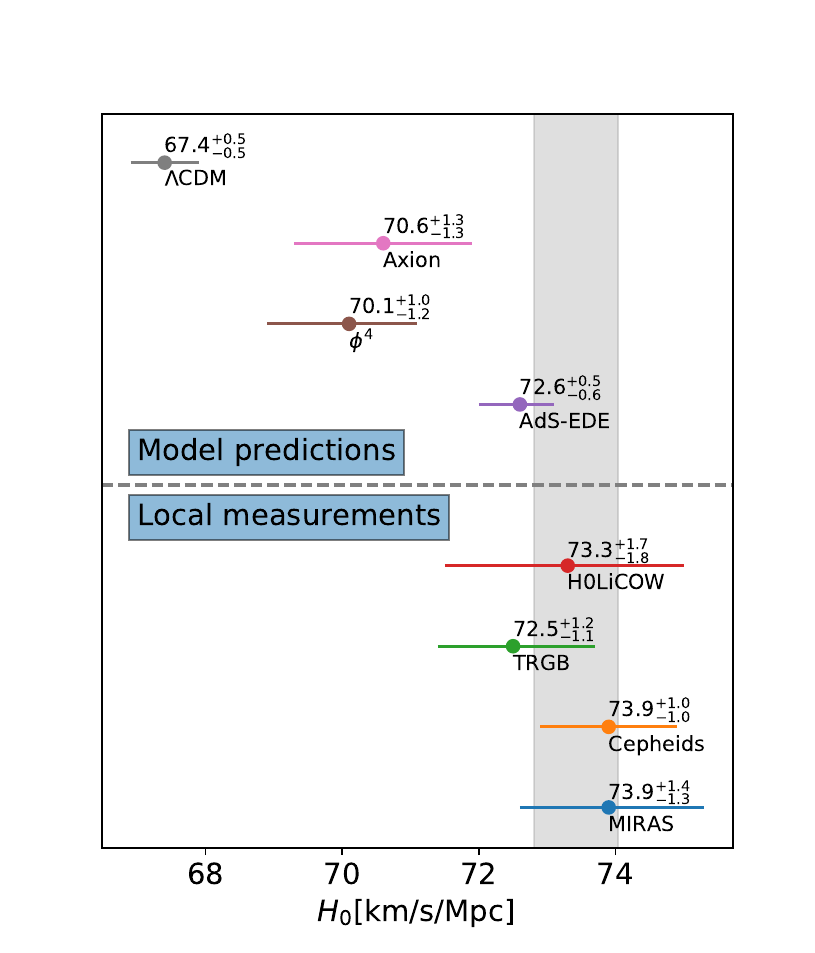}
\caption{Predicted $H_0$ values in the $\Lambda$CDM, axion-like \cite{Poulin:2018cxd}, $\phi^4$ \cite{Agrawal:2019lmo} and AdS-EDE models confronted with various mutually independent
local measurements. MIRAS, Cepheids and TRGB stands for three
independent Type Ia supernova calibrators in the local distance
ladder (see Ref.\cite{Verde:2019ivm}). The $H_0$
measured by lensing time delays from lensed quasars, independent
of the distance ladder approach, is also included (the H0LiCOW
result \cite{Wong:2019kwg}). Gray band represents the combined
$1\sigma$ region of the measured $H_0$.} \label{H0}
\end{figure}

We, based on Eqs.\eqref{CMB}, \eqref{BAO-trans} and
\eqref{BAO-rad} (i.e. CMB and BAO constraints), identified why EDE
is compatible with $T_{0,FIRAS}$, while lifting $H_0$ at the cost of an enhanced $\omega_m$. As a contrast,
we also showed that $w$CDM models with $w\lesssim -1.3$ seem
inconsistent with $T_{0,FIRAS}$. We performed MCMC analysis for
the corresponding models to confirm our observations. It has been
pointed out in Ref.\cite{Ivanov:2020mfr} that for $\Lambda$CDM,
$T_0$ yielded by the Planck and SH0ES data has $>4\sigma$
discrepancy compared with $T_{0,FIRAS}$. However, we showed that
EDE is compatible with not only $T_{0,FIRAS}$, but also known independent local measurements
of $H_0$, see Fig-\ref{H0} and Appendix-\ref{apx:strong lensing}. As argued in
section-\ref{sec:constraint}, $H_0$ is lifted at the cost of an
enlarged $\omega_m$ and also an enhanced $n_s$ compensating for diffusion damping at high $l$, the
corresponding models are thus expected to receive tight
constraints with the inclusion of matter power spectrum data
\cite{Hill:2020osr,Ivanov:2020ril,DAmico:2020ods}, which will be
left for future study. Still, our result suggests that even if
EDE is not the final story restoring cosmological concordance, it
might be on the right road and relevant issues are worth studying.

\begin{figure}[h]
    \centering
    \includegraphics[width=0.8\linewidth]{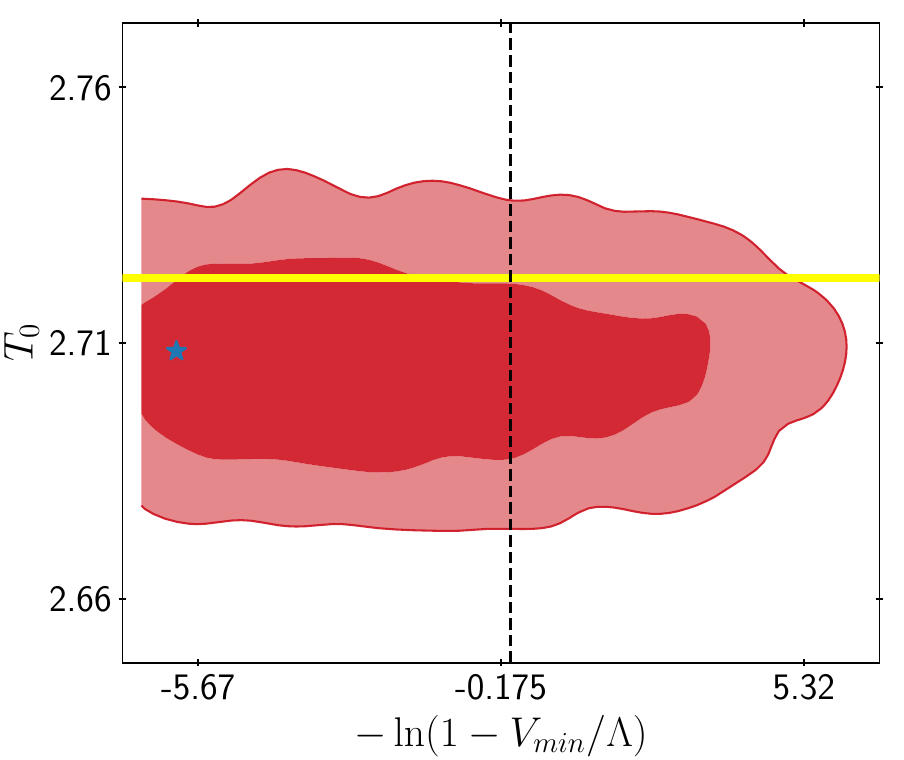}
    \caption{Marginalized contour of $T_0$ with respect to $V_{min}/\Lambda$.
     The axis $-\ln(1-V_{min}/\Lambda)$ is chosen such that it is log scale when
$-V_{min}/\Lambda\ll1$ (deep in the AdS phase) and
$V_{min}/\Lambda\to1$, while it is linear around $V_{min}\sim0$.
Dashed line labels $V_{min}=0$. Yellow band represents
$T_{0,FIRAS}$.}
    \label{T0-AdS}
\end{figure}

Inspired by the $\alpha$-attractor
\cite{Carrasco:2015rva,Akrami:2017cir}, we also presented a
well-motivated AdS-EDE model. In the MCMC analysis, we do not
assume AdS \textit{in priori}, but in Fig-\ref{T0-AdS} we see that the MCMC
result weakly hints the existence of an AdS phase, with the
best-fit cosmology having AdS depth $V_{min}\sim -(0.001$eV$)^4$. The best-fit
model allows $H_0\sim 73$km/s/Mpc at 1$\sigma$ range, which
indicates that the existence of AdS phase around recombination
helps to significantly lift $H_0$. Our result again highlights an
unexpected point that AdS vacua, ubiquitous in consistent
UV-complete theories, might also play a crucial role in our
observable Universe.

\paragraph*{Acknowledgments}

This work is supported by the University of Chinese Academy of
Sciences. Y.S.P. is supported by NSFC, Nos. 11575188, 11690021.
The computations are performed on the TianHe-II supercomputer.

\appendix
\section{MCMC results of the $\alpha$AdS model}\label{apx:alpha-attractor}
\begin{figure}
\centering
\includegraphics[width=\linewidth]{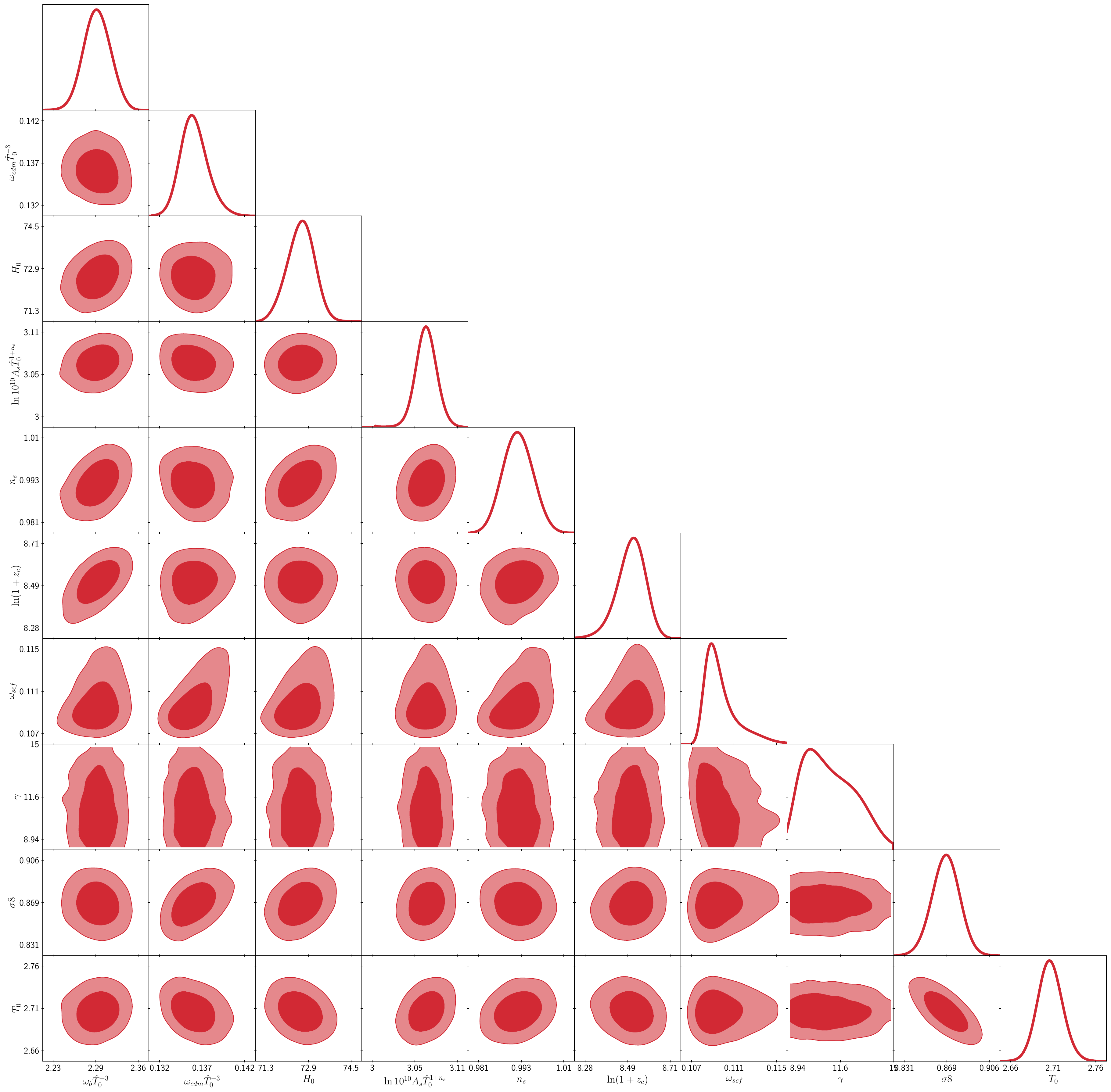}
\caption{Posterior distributions and marginalized $68\%$ and
$95\%$ contours of all model parameters in the $\alpha$AdS model
confronted with the full datasets P18+BAO+SN+$H_0$.}
\label{sg triangle}
\end{figure}

In the MCMC analysis we sample over $\{\omega_b/\hat{T}_0^3,
\omega_{cdm}/\hat{T}_0^3, H_0,
\ln(10^{10}A_s\hat{T}_0^{1+n_s}),n_s,\tau_{reio},$
$T_0,\omega_{scf},\ln(1+z_c),\gamma\}$. We use flat priors for additional EDE
parameters (Table-\ref{prior table}). Here, we do not assume AdS \textit{in priori} in the
MCMC analysis, since the $\gamma$ prior in Table-\ref{prior table}
covers non-AdS region of the potential, see
Eq.\eqref{gamma13}. Posterior distributions and marginalized
contours of all cosmological parameters are plotted in Fig-\ref{sg triangle}.
The mean and best-fit values are shown in Table-\ref{sg
value}. We also report the best-fit $\chi^2$ values per experiment
in Table-\ref{sg bestfit}.

\begin{table}
    \centering
    \caption{Flat priors of $\alpha$AdS parameters}
    \begin{tabular}{|c|c|}
        \hline
        &prior\\
        \hline
        $\omega_{scf}$&$[10^{-4},0.4]$\\
        \hline
        $\ln(1+z_c)$&$[7.5,9.5]$\\
        \hline
        $\gamma$&$[5,15]$\\
        \hline
    \end{tabular}
    \label{prior table}
\end{table}
\begin{table}[h]
    \caption{Mean and best-fit values of all model parameters}
    \begin{tabular}{|l|c|c|c|c|}
        \hline
        Param & best-fit & mean$\pm\sigma$ & 95\% lower & 95\% upper \\ \hline
        $100\omega_b \hat{T}_0^{-3}$ &$2.278$ & $2.273_{-0.047}^{+0.044}$ & $2.182$ & $2.365$ \\
        $\omega_{cdm} \hat{T}_0^{-3}$ &$0.1333$ & $0.1345_{-0.0025}^{+0.002}$ & $0.1302$ & $0.1391$ \\
        $H_{0 }$ &$72.54$ & $72.57_{-0.53}^{+0.52}$ & $71.56$ & $73.63$ \\
        $\ln(10^{10}A_s \hat{T}_0^{1+n_s})$ &$3.076$ & $3.077_{-0.015}^{+0.016}$ & $3.046$ & $3.108$ \\
        $n_{s }$ &$0.9939$ & $0.9926_{-0.0044}^{+0.0043}$ & $0.9839$ & $1.001$ \\
        $\ln (1+z_c )$ &$8.479$ & $8.51_{-0.061}^{+0.076}$ & $8.362$ & $8.651$ \\
        $\omega_{scf }$ &$0.1091$ & $0.1098_{-0.002}^{+0.0006}$ & $0.1073$ & $0.1134$ \\
        $\gamma$ &$8.748$ & $10.98_{-2.2}^{+1}$ & $8.444$ & $13.92$ \\
        $T_{0 }$ &$2.711$ & $2.709_{-0.016}^{+0.015}$ & $2.679$ & $2.74$ \\
        $\sigma8$ &$0.8635$ & $0.8677_{-0.012}^{+0.012}$ & $0.8435$ & $0.8922$ \\
        \hline
    \end{tabular}
    \label{sg value}
\end{table}

\begin{table}
\caption{best-fit $\chi^2$ per experiment}
\begin{tabular}{|c|c|}
\hline
Experiment&$\chi^2$\\
\hline
Planck high $l$&$2347.44$\\
Planck low $l$&$416.89$\\
Planck lensing&$11.79$\\
BAO BOSS DR12&$0.66$\\
BAO low $z$&$2.46$\\
Pantheon&$1026.94$\\
SH0ES&$1.33$\\
\hline
\end{tabular}
\label{sg bestfit}
\end{table}

\section{MCMC results including the lensing time delay measurements}\label{apx:strong lensing}
\begin{figure}[h]
\centering
\includegraphics[width=0.8\linewidth]{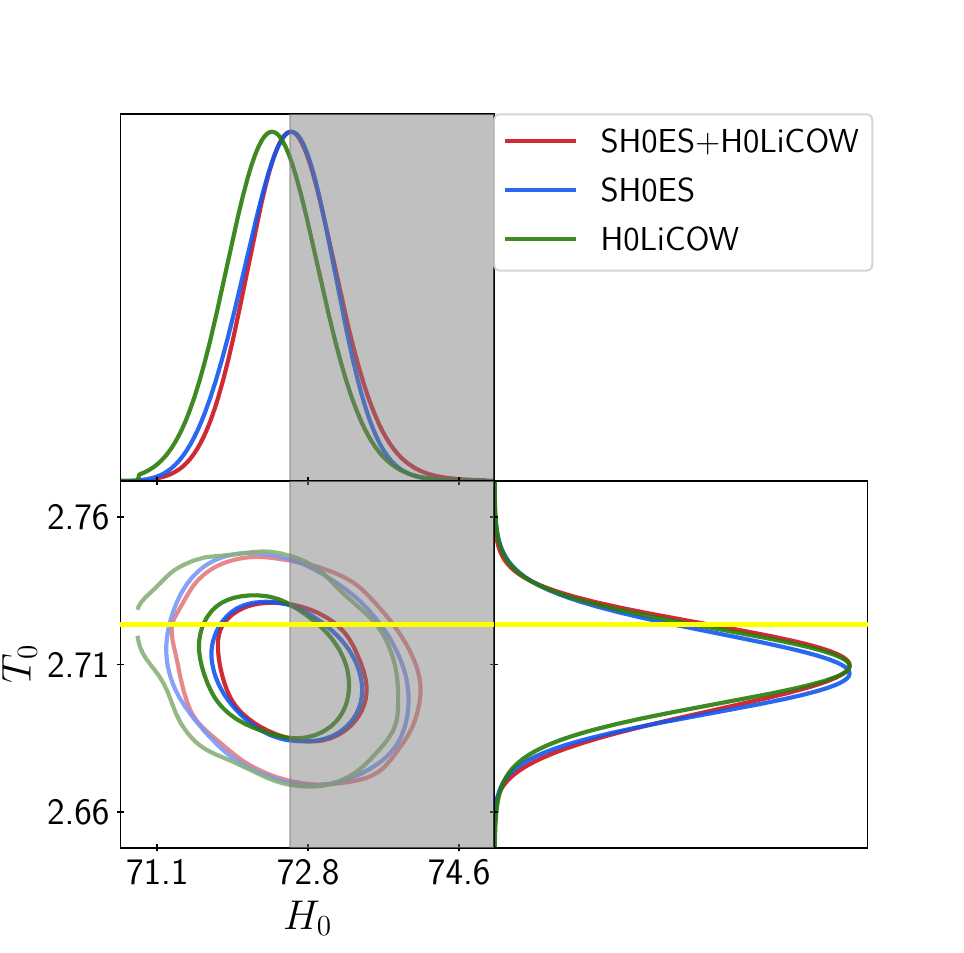}
\caption{Posterior distributions and 68\% and 95\% contours of $T_0$ and $H_0$ in the $\alpha$AdS model confronted with P18+BAO+SN+SH0ES/H0LiCOW and P18+BAO+SN+SH0ES+H0LiCOW. The gray (yellow) band corresponds to the 1$\sigma$ region of the SH0ES (FIRAS) measurement.}
\label{T0-H0-lens}
\end{figure}
\begin{figure}
\centering
\includegraphics[width=0.8\linewidth]{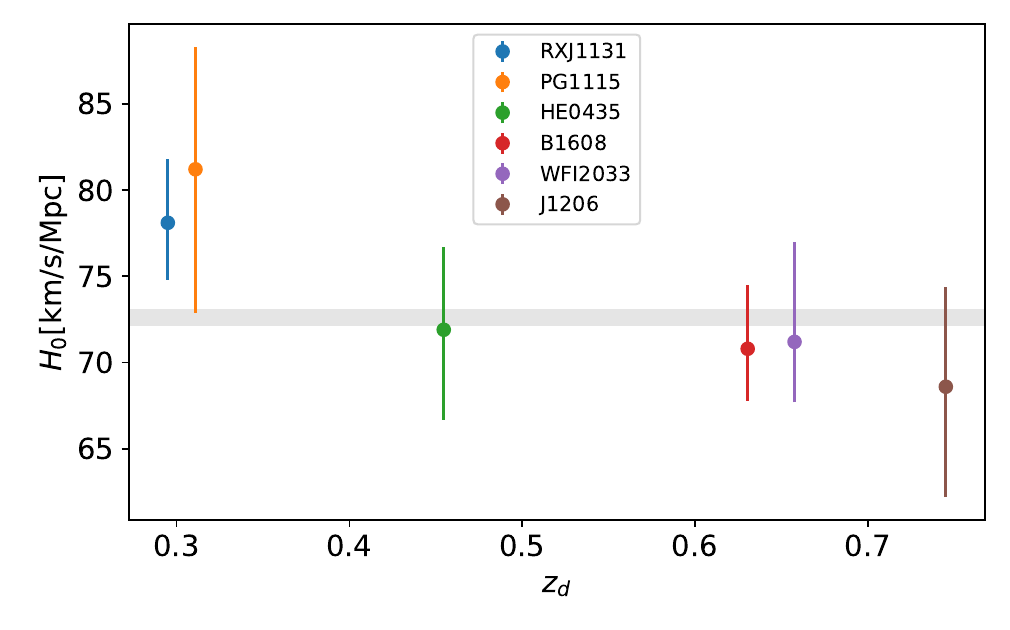}
\caption{Measured values of $H_0$ per lens of H0LiCOW versus lens redshift $z_d$ in the best-fit $\alpha$AdS model. The lenses presented are B1608 \cite{Suyu:2010ApJ,Jee:2019hah}, HE0435 \cite{Wong:2016dpo,Chen:2019ejq}, J1206 \cite{Birrer:2018vtm}, RXJ1131 \cite{Chen:2019ejq,Suyu:2013kha}, PG1115 \cite{Chen:2019ejq} and WFI2033 \cite{Rusu:2019xrq}.}
\label{H0-lens}
\end{figure}
All MCMC analyses presented in the main text employ the SH0ES result as a Gaussian prior, it is thus worth checking if the results are compatible with other independent experiments. The current $H_0$ can be measured by lensing time delays of strongly lensed quasar images, completely independent of the local distance ladder approach, in particular SH0ES. To this end we re-analyze the $\alpha$AdS model \eqref{sg potential} with the same ten model parameters as discussed in the main text and the dataset P18+BAO+SN+H0LiCOW. In Fig-\ref{T0-H0-lens} we plot relevant $T_0-H_0$ contours and posterior distributions. We also include a combining all (H0LiCOW+SH0ES) result since the two datasets are independent of each other. It is clear that the MCMC analysis with the full H0LiCOW likelihood code yields nearly identical results to that with the SH0ES prior, confirming the robustness of treating local $H_0$ measurements as simple Gaussian priors in the MCMC analysis\footnote{The mean $H_0$ predicted by the P18+BAO+SN+H0LiCOW chain is slightly smaller than that of the SH0ES chain. The SH0ES prior also seems to dominate over the H0LiCOW data in the combining all chain. We attribute this to the larger mean value and shorter error bar of $H_0$ in the SH0ES measurement.}. Additionally, we vary $H_0$, for each of the six lenses in H0LiCOW, with all other parameters fixed at their best-fit values in $\alpha$AdS and plot the results in Fig-\ref{H0-lens}. Ref.\cite{Wong:2019kwg} noted a possible trend (not yet statistically significant due to the small sample size) of lower lens redshift systems having a larger inferred value of $H_0$. Fig-\ref{H0-lens} shows such a trend, if really  exists, seems to persist in the AdS-EDE model. 

\section{Scalar field perturbations in EDE and $\omega_m$}\label{apx:omegam in ede}
\begin{figure}[h]
	\centering
	\includegraphics[width=0.8\linewidth]{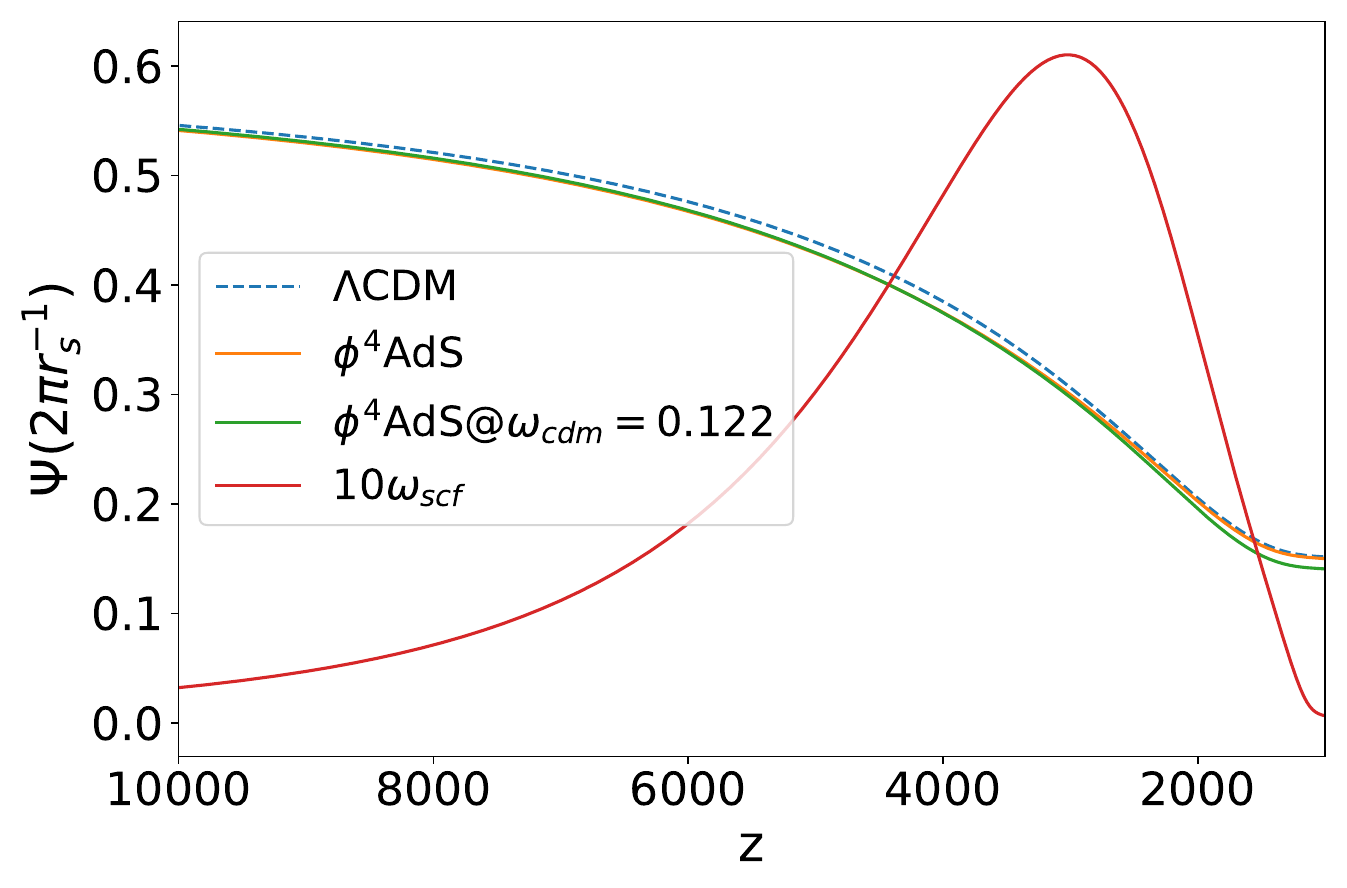}
	\caption{Evolution of $\Psi$ with $k=2\pi/r_s$, which roughly
		corresponds to the first acoustic peak, plotted for the best-fit
		models of $\Lambda$CDM and $\phi^4$AdS. The green line is produced by
		a $\phi^4$AdS model with reduced $\omega_{cdm}$ while fixing all other
		parameters to the best-fit.}
	\label{Psi}
\end{figure}

When the EDE becomes non-negligible, the gravitational
perturbation $\Psi$ will be suppressed by the EDE perturbations
\cite{Hill:2020osr}. In order to preserve the fit to the CMB data,
$\omega_m$ must increase accordingly to compensate for the slight
suppress in $\Psi$.

We plot the evolution of $\Psi$ in Fig-\ref{Psi}. Two EDE lines
are nearly identical at high-$z$ due to the same cosmological
parameters except for $\omega_{cdm}$. However, they will not coincide
any longer when EDE becomes non-negligible. $\Psi$ in the $\phi^4$AdS
model with fixed $\omega_{cdm}=0.122$ is suppressed compared with that
in the best-fit $\phi^4$AdS model. This is because in the best-fit
$\phi^4$AdS model such suppression will be compensated by the
gravity of extra dark matter abundance, which lifts $\Psi$ at
recombination to the $\Lambda$CDM value (dashed line), so produces
correct power in the CMB TT spectrum.

\end{document}